\documentstyle [12pt] {article}

\catcode`@=12
\begin{document}
\thispagestyle{empty}
{}
\vspace{0.5in}

\begin{center}
{\Large \bf Constraints on background torsion field from K physics}

\vspace{0.5in}

{\bf Subhendra Mohanty$^{a)}$ and Utpal Sarkar$^{a),b)}$\\}
\vspace{0.3in}
$^{a)}$ {\sl Theory Group, Physical Research Laboratory,
Ahmedabad, 380 009, India}\\
$^{b)}$ {\sl Abdus Salam International Center for Theoretical Physics, 
34100 Trieste, Italy}\\

\vspace{1.0in}

\end{center}

\begin{abstract}\

We point out that a background torsion field will produce
an effective potential to the $K$ and $\bar{K}$ with opposite signs. 
This  allows us to constrain the background torsion field from
the $K_L$ and $K_S$ mass difference, $CPT$ violating 
$K^\circ$ and $\bar{K^\circ}$ mass difference and the $CP$ violating 
quantities $\epsilon$ and $\eta_{+-}$. The most stringent bound 
on the cosmological background torsion  $\langle T^0 \rangle < 
10^{-25}$ GeV comes from the direct measurement of the $CPT$ violation.

\end{abstract}

\newpage
\baselineskip 24pt

General Theory of relativity has so far succeeded in confronting all 
experimental tests. However the problem of quantising gravity
leads one
to believe that Einsteins theory though correct may not be the most
general theory which describes the dynamics of the metric tensor and its
interactions with matter. The ultimate quantum theory of gravity must
also explain the low energy phenomenology. This gave rise to the birth
of string theory, which is now considered as the most consistent theory
of quantum gravity. In string theory the metric tensor field comes out
naturally and gives the response of matter to this metric. However, it
also predicts several other fields like the antisymmetric second rank
field, which enters via its antisymmetrized derivatives, $T_{\alpha
\beta \gamma} = \partial_{[\alpha} A_{\beta \gamma ]}$, which are usually
referred to as the torsion field. In addition to the string inspired 
approach to study the torsion dynamics, there are some modifications of
the GTR where connection is treated as more fundamental than the metric.

In a metric compatible theory of gravity one generalizes the connection
by including the torsion tensor \cite{books,carol}. The symmetric part of
the generalized connection  
are the Christoffel symbols given by the usual formula in terms of the
metric,
whereas the torsion is the antisymmetric part of the connection which in
Einsteins gravity is assumed to be zero.
The coupling of the matter fields to torsion arises from the covariant
derivative with respect to the generalized connection . The covariant
derivative of a fundamental scalar field is  just the partial derivative
therefore the torsion term does not couple to scalars .
The torsion field also does not couple to
electro-magnetic fields as the gauge invariant field strength is the 
antisymmetric partial derivative,$F_{\mu\nu}\equiv ({\rm d}A)_{\mu\nu}
  =\partial_\mu A_\nu -\partial_\nu A_\mu\ $. This is the general
definition for the field strength even in curved space. If one were to 
generalise the definition of $F_{\mu\nu}$ by replacing the partial
derivatives with the covariant derivative (\ref{cov}), the extra terms
involving torsion will not be gauge invariant. Coupling of the
electro-magnetic field with torsion can only arise in some more generalized
versions of torsion theories \cite{parera}. 

It has long been known that fermions can couple with torsion field
\cite{books,carol}. The
fermions couple as an axial current to the dual of the torsion tensor.
 Since an axial current is spin dependent in the forward
scattering limit, the torsion force on a macroscopic collection of
fermions will average to zero unless the fermion spins is polarized (which
is difficult to attain in experiments). For this reason one cannot
constrain the torsion couplings from the  usual fifth force experiments
\cite{will}.

In this paper we point out that composite scalars like mesons can couple
to the torsion field through  the constituent quarks. In a pseudoscalar
meson the torsion
potential being spin dependent couples to the difference of the dipole 
moments of the quarks .  We show that in a
background torsion field , the potentials of the $K^\circ$ and $\bar K^0$
have  opposite signs. This apparent violation of CPT
and CP in the $K-\bar K$ system can be constrained from the
 kaon oscillation
experiments and these constraints allow us to put bounds on the 
 cosmological torsion background to be $\langle T^0 \rangle < 10^{-25}$ GeV .

We start with a brief review of gravity with torsion and then point out 
how the torsion field couple to the K system. We then parameterize the 
effective potential due to this non vanishing torsion field and put bound
from various experiments. 

In a metric compatible theory of gravity one generalise s the connection
by including the torsion tensor \cite{books,carol} ,
\begin{equation}
\Gamma^\alpha_{\mu\nu}\equiv\left\{\matrix{\alpha\cr
  \mu\nu\cr}\right\}+{1\over 2}\left(T_{\mu\nu}{}^\alpha  
  -T_\nu{}^\alpha{}_\mu
  +T^\alpha{}_{\mu\nu}\right)
\label{one}
\end{equation}
where the Christoffel symbols $\left\{\matrix{\alpha\cr 
\mu\nu\cr}\right\}$ are the symmetric part of $\Gamma^\alpha_{\mu\nu} $ 
and are given by the usual formula
\begin{equation}
\left\{\matrix{\alpha\cr \mu\nu\cr}\right\}
  \equiv{1\over 2}g^{\alpha\beta}\left(\partial_\mu g_{\nu\beta}
  +\partial_\nu g_{\beta\mu}-\partial_\beta g_{\mu\nu}\right).
\end{equation}
whereas the torsion is the antisymmetric part of $\Gamma^\alpha_{\mu\nu} $,
\begin{equation}
{T_{\rho \sigma}}^\mu = \Gamma^\mu_{\rho \sigma} -\Gamma^\mu_{\sigma \rho} 
\label{T}
\end{equation}
The coupling of the matter fields to torsion arises from the covariant
derivative with respect to the generalised connection (\ref{one}), 
\begin{equation}
\nabla_\mu X^\alpha \equiv \partial_\mu X^\alpha +  
\Gamma^\alpha_{\mu\nu} X^\nu
\label{cov}
\end{equation}
The minimal coupling action of the Dirac spinor fields in an external
gravitational 
field with torsion will now become
\begin{equation}
{\cal L} = {i \over 2} \int d^4 x \left( \nabla_\mu \bar{\psi}
\gamma^\mu \psi -
\bar{\psi} \gamma^\mu \nabla_\mu \psi - 2 i m \bar{\psi} \psi \right) .
\end{equation}
This contains the usual torsion free part of the lagrangian and the
interaction part is given by (we set $g_{\mu \nu } = \eta_{\mu \nu}$),
\begin{equation}
{\cal L}_{I} = {i \over 8} \int d^4 x T_{\mu \nu \lambda} \bar{\psi} 
\gamma^{[\mu} \gamma^\nu \gamma^{\lambda ]} \psi = {3 \over 4} T^\sigma
(\bar{\psi} \gamma_5 \gamma_\sigma \psi).
\end{equation}
where the pseudo-trace irreducible component of the torsion field is
defined as dual to the antisymmetric part of the torsion field,
\begin{equation}
 T^\sigma \equiv {1 \over 3!} \epsilon^{\mu \nu \lambda \sigma}
T_{\mu \nu \lambda} .
\end{equation}
 Thus the torsion coupling to matter reduces to 
a coupling of an axial vector current to a torsion pseudo-vector
$T_\sigma$.
The other components of the torsion, namely, the trace $T_\beta =
T^\alpha_{\phantom{\alpha} \beta \alpha} $ and the tensor $q^\alpha_{
\phantom{\alpha} \beta \lambda} $ couples to this pseudo-vector 
component but not with matter directly. From the fact that the torsion
pseudo-vector $T^\sigma$ couples to a axial vector current , one can assign
the following transformation  properties  to $T^\sigma= (T^0,\vec T)$
under transformations of 
 Charge conjugation (C), Parity (P) and Time reversal (T) symmetries :
\begin{equation}
\matrix{ &C &P&T\cr
(T^0, \vec T)\Rightarrow&(T^0, \vec T)&(-T^0, \vec T)&(T^0, -\vec T)}
\end{equation}
and under the combined operation of $CPT : T^{\sigma} \rightarrow
-T^{\sigma}$.
 We shall now assume that there is a fixed background potential
due to a non vanishing value of a background  torsion, which
breaks
$CP$ ,  $T$ and $CPT$ along with the local Lorentz invariance . However,
we would
like to preserve the isotropy of the field and assume a non-vanishing 
value only time-like component of the pseudo-vector $(3/4)<T^0> = t^0$,
where
$t^0$
has a dimension of mass. In this way a non-zero  background $t$ breaks CP
and CPT,
and this should give rise to some observable effects of these symmetry
violations.

In the rest of this article we study if it is
possible to constrain this background potential due to $t$ from some
experiments. We point out that the background potential couples to $K$ and
$\bar{K}$ in a different way, and we get an additional
$CP$ and $CPT$ violating potential for the K-system. 
Since the torsion term couples to fermions as a axial vector, in the forward
scattering limit  the dipole moment of the fermion gives the dominant
coupling. This can be seen by writing the Gordon decomposition of
the fermion axial current as
\begin{equation}
\bar \psi( p^{\prime})\gamma_{\mu} \gamma_5 \psi(p)=\bar \psi(p^{\prime})
\left[{(p^{\prime} - p)_{\mu} \over 2m} + {i\over 2m} \sigma_{\mu
\nu}~(p^{\prime}
+p )^{\nu} \gamma_5 \right]\psi(p)
\label{gor}
\end{equation}
In the forward scattering limit the effective potential of a fermion due 
to a background torsion is
\begin{equation}
{t^0\over m}~ \bar \psi(p) ~\sigma_{0 i}~\gamma_5 p^i ~\psi(p)= {2t^0\over
m}~\vec s\cdot \vec p = {2t^0~ |\vec p| \over m}~ \lambda
\label{fpot}
\end{equation}
where $\vec s$ is the spin and $\lambda$ is the helicity of the fermion.
At high energies the ($E>>m$) the helicity is a conserved (upto
$O(m^2/E^2)$) and one can assign a fermion in   a torsion background $t$
the potential $ \pm t^0 |\vec p| /m$ where the sign depends upon the 
helicity. Unlike the gravitational potential energy, the torsion potential
energy of a macroscopic body is not large as the spins of a macroscopic
body are aligned randomly. Due to this reason it is not
possible to put bounds on the torsion  force with any of the usual
fifth force experiments \cite{will}. 
In a psedoscalar meson like $K^\circ $ or  $\bar K^\circ $ the  quark and
the
anti-quark have
opposite spins and at high energies they can be assigned the helicities 
$1/2$ and $-1/2$ respectively. The net dipole form factor of mesons will
be non-zero if the quark and anti-quark have different masses. We can write
the quark model of the $K^\circ$ as 
\begin{equation}
|K \rangle = {1\over \sqrt 2}[ s(+)~ \bar d(-) - s(-) ~\bar d(+)]
\label{K}
\end{equation}
The signs in the brackets indicate the helicity quantum numbers of the
s-quark and the d-anti-quark. The combination appears with a relative
negative sign as under Parity the helicity flips sign and the
mesons being pseudo-scalars under Parity $K^\circ \rightarrow \bar
K^\circ$.
The net torsion potential of the quark-antiquark combination (\ref{K}) of
the $K^\circ$ meson is
\begin{equation}
 t^0 {|\vec p|\over \sqrt 2} [ ({1\over m_s} - {1\over m_d})- 
({-1\over m_s} - {1\over m_d})) 
=  - t^0 |\vec p| {\sqrt 2} {(m_s-m_d)\over m_s m_d}
\label{VK}
\end{equation}
The quark quantum numbers of the $\bar K^\circ$ is given by
\begin{equation}
|\bar K \rangle = {1\over \sqrt 2}[ d(+)~ \bar s(-) - d(-)~ \bar s(+)]
\label{Kbar}
\end{equation}
This is consistent with (\ref{K}) and the requirement that under CP :
$|K^\circ\rangle
\rightarrow -|\bar K^\circ\rangle$. The net torsion potential of the $\bar
K^\circ$ meson is therefore
\begin{equation}
 t^0 {|\vec p|\over \sqrt 2} [ ({1\over m_d} - {1\over m_s})- 
({-1\over m_d} -
 {1\over m_s})] =  t^0 |\vec p| {\sqrt 2}  {(m_s-m_d)\over m_s m_d}
\label{VKbar} 
\end{equation}
So we can see from the torsion potentials (\ref{VK}) for $K^\circ$ and
(\ref{VKbar}) for $\bar K^\circ$ that the two terms have opposite sign
as is 
expected from the fact that when the background torsion field is non-zero
then the potential must be $CP$ and $CPT$ violating.

In the basis [$K^\circ~~~\bar{K^\circ}$] the background potential
due to the non-vanishing torsion field now becomes
\begin{equation}
{\cal V} = -~V~~\pmatrix{1 & 0 \cr 0 & -1} 
\end{equation}
where $V = {t \over m} |\vec{p}|$; $m$ is the kaon mass and 
$t \equiv t^0 {\sqrt 2} (m_s - m_u)m_K /( m_u m_s)$ is the background 
torsion parameter.

 We shall be comparing our results with 
experiments where the kaons are ultra-relativistic. 
The total hamiltonian in the basis [$K^\circ~~~~\bar{K^\circ}$] 
will now become,
\begin{eqnarray}
H &=&
p I + {1 \over 2 p} \pmatrix{m - {i \over 2} \Gamma
& {1 \over 2} \left(\delta m - {i \over 2} \delta \Gamma \right)
\cr {1 \over 2} \left(\delta m - {i \over 2} \delta \Gamma \right)
& m - {i \over 2} \Gamma}^2 + {\cal V} \nonumber\\
&\equiv& p I + {1 \over 2 p} \pmatrix{M_+ - {i \over 2} \Gamma_+
& M_{12} - {i \over 2} \Gamma_{12} \cr M_{21} - {i \over 2} 
\Gamma_{21} & M_- - {i \over 2} \Gamma_-}^2 
\end{eqnarray}
where $I$ is the identity matrix. If $CPT$ is conserved, then $M_+ = M_-$
and $\Gamma_+ = \Gamma_-$. The direct $CPT$ violating quantity $M_+ - M_-$
is then given by,
\begin{equation}
M_+^2 - M_-^2 = 4 p V \hskip .25in {\rm or} \hskip .25in
|M_+ - M_-| = 2 \left(p\over m \right)^2 t,
\end{equation}
which depends quadratically on energy. Similar 
energy dependence in the $K$ system was discussed
in the literature \cite{lee,nacht,hambye,Kenyon}, 
which could arise from violation of the weak equivalence
principle or the violation of the Local Lorentz invariance.

We shall first constrain the background 
torsion $T$ from the measurements of the $K_L$ and $K_S$ mass difference, 
the comparison of the induced 
$CP$ violation due to the $CPT$ violation, and from
the direct measurement of the $CPT$ violating quantity -- 
the mass difference of $K^\circ$ and $\bar{K^\circ}$.

In the basis of the physical states [$K_L~~~K_S$] the hamiltonian will
become,
\begin{equation}
H =  \pmatrix{p + {m_L^2 \over 2 p} & 0 \cr 0  & p + {m_S^2 \over 2 p}}.
\end{equation}
Comparing the two we can write down the masses of these physical states
$K_L$ and $K_S$ as,
\begin{eqnarray}
m_L^2 &=& [(m+ {\delta m\over 2})-{i\over2}( \Gamma
+{\delta \Gamma \over 2})]^2
 + {2 p^2 V^2\over 
(m - {i \over 2} \Gamma)(\delta m - {i \over 2} \delta \Gamma)} \nonumber\\[8pt]
m_S^2 &=&  [(m - {\delta m\over 2})-{i\over2}( \Gamma
 - {\delta \Gamma \over 2})]^2
 - {2 p^2 V^2\over 
(m - {i \over 2} \Gamma)(\delta m - {i \over 2} \delta \Gamma)}
\label{eigs}
\end{eqnarray}

To constrain the torsion parameter we now write down the 
$K_L$ and $K_S$ mass difference as,
\begin{equation}
m_L - m_S = \left[ (\delta m)^2 + 4 \left( { p^2 \over m^2 } t \right)^2 
\right]^{1/2}
\end{equation}
For the experimental value of the  $K_L$ and $K_S$ mass difference we 
consider the CDF experiments \cite{gib,sch}, 
which were done at energies as high as 
160 GeV. Although earlier low energy experiments \cite{cul,gew,gje}
differ from these 
CDF values by 2 $\sigma$, a recent low energy experiment at 
CERN \cite{adl} gives a
value close to the CDF values. Another advantage of CDF value is that
it is done at higher energies. So because of the energy dependence of
the torsion parameter, the bound will be stronger.
Taking the experimental value of the $K_L$ and $K_S$ mass difference 
to be $(m_L - m_S)_{expt} = (.528 \pm .0030) \times 10^{10} \hbar 
s^{-1} = 3.49 \times 10^{-15} GeV$ 
we get an bound on the torsion parameter to be, 
$t < 1.3 \times 10^{-20} $ GeV. 

Usually the bound on the CPT violating parameters is obtained from
a direct measurement \cite{car} of the upper bound on $|M_+ - M_-|/m_K$, 
which is $9 \times 10^{-19}$ \cite{pdg}. Although the CPLEAR bound on
the direct measurement on the bound of the $K^\circ$ and $\bar{K^\circ}$ 
mass difference is stronger than the NA31, we use the latter bound
because of the energy dependence of the torsion parameter. The bound
on the former would constrain strongly the amount of CPT violation arising from 
the string motivated violation of quantum mechanics \cite{ellis,cpl}.
The NA31 experiment was 
done at energies around 100 GeV \cite{car}, which gives the strongest 
bound on the torsion parameter $$t =
{1 \over 2} \left( {m \over p}\right)^2 |M_+ - M_-| 
< 5.6 \times 10^{-24} GeV.$$

Another approach of constraining the CPT violating parameter is by  
following Kenyon \cite{Kenyon}. They introduce the parameter,
\begin{equation}
\Delta = {1 \over 2} {M_+ - M_- \over \delta M - \delta \Gamma},
\end{equation}
where $\delta M = m_L - m_S$ and $\delta \Gamma = \Gamma_L - \Gamma_S$.
They relate this parameter to the CPT violating parameter, in our case 
it will be the torsion parameter, as
$$  \left({p \over m}\right)^2 t = 2 \Delta M Im (\Delta). $$
Using the Bell-Steinberger relation they obtain, 
$Im(\Delta) \leq 2 \times 10^{-4}$. This implies a bound on the 
torsion parameter, $t< 1.75 \times 10^{-23}$ GeV. 

We shall now constrain the torsion parameter from an analysis of the 
measurements of the $CP$ violating quantities. If we assume that 
the observed $CP$ violation comes entirely from the torsion parameter,
then we are led to immediate contradiction, because of two reasons,
as we shall discuss next. However, it is possible to constrain
the torsion parameter from the measurement of the $CP$ violating 
quantities in the $K$ system. 

The eigenfunctions whose time evolution is given by
$|K_L (t) \rangle = |K_L \rangle~~~exp\{-i m_L t\}$ and
$|K_S (t) \rangle = |K_S \rangle~~~exp\{-i m_S t\}$ are given by the
expressions
\begin{eqnarray}
|K_L \rangle = {1\over (2(1+|\epsilon|^2)^{1/2}} \left( (1+\epsilon)
|K^\circ \rangle - (1 -\epsilon)|\bar K^\circ \rangle \right)
\nonumber\\[8pt]
|K_S \rangle = {1\over (2(1+|\epsilon|^2)^{1/2}} \left( (1+\epsilon)
|K^\circ \rangle + (1 -\epsilon)|\bar K^\circ \rangle \right)
\label{eifs}
\end{eqnarray}
where the mixing parameter $\epsilon$ is given by
\begin{equation}
\epsilon ={ 2 p V \over (m - {i \over 2} \Gamma)
(\delta m - {i \over 2} \delta \Gamma)}
\label{epsi}
\end{equation}
 From (\ref{eifs}) we find that the  mixing parameter between $ |K_L
\rangle $ and $|K_S \rangle $ is given by
\begin{equation}
\langle K_S |  K_L \rangle = 2 Re~ \epsilon  = 4 \left( p \over m
\right)^2
~t~~{\delta m \over (\delta m)^2 + ({\delta \Gamma \over 2})^2}
\end{equation}
In this derivation we have followed the possibility that the  mixing
between $K_L$ and 
$K_S$ - the source of $CP$ violation
is only due to the torsion potential. This assumption implies $\eta_{+-} =
\epsilon$
and taking the experimental values \cite{pdg} of 
${\delta \Gamma \over 2} = 
3.68 \times 10^{-15}$ GeV; it predicts $\phi_{+-} \approx 45^\circ$
as in the super-weak model which is ruled out experimentally. This and the
non-observation of the energy
dependence of $\eta_{+-}$ rules out the possibility of explaining the
observed
$CP$ violation in kaons entirely from the torsion background. However, if
we assume 
that the constant value of $\eta_{+-}$ till the highest energy of the 
CDF experiments (160 GeV) is not due to the torsion parameter then we 
can put bound on this parameter. Taking ${\rm Re} 
\epsilon = 2.27 \times 10^{-3}$, we get,
\begin{equation}
t = {1\over 2} (Re~ \epsilon) ({m\over p})^2~~~ {(\delta m)^2 +
({\delta \Gamma \over 2})^2 \over \delta m}
< 8.5 \times 10^{-23} GeV .
\end{equation}
In other words, if we consider the bound on the torsion parameter 
from the direct measurement of the $CPT$ violating $K^\circ$ and 
$\bar K^\circ$ mass difference, then the prediction for the $CP$
violation would be less than what has been observed experimentally.

The  most stringent bound $t <  5.6 \times 10^{-24} GeV$ therefore arises
from the NA31 constraint on $|M_+ - M_-|$ leads to the following bound
on the time-like component of the torsion pseudo-vector $T^\sigma$,
\begin{equation}
\langle T^0 \rangle = {4\over 3} t^0 = {4\over 3 \sqrt 2}~{m_u m_s
\over (m_s-m_u) m_K} t < 10^{-25} ~~GeV
\end{equation}

To summarize, we pointed out that if there is any background torsion 
field, it will lead to an apparent violation of CP and CPT in composite
pseudo-scalar particles.
As a result one can constrain this parameter severely from the K 
system. The measurements of the direct $CPT$ violating parameter, which
is the mass difference of the $K^\circ$ and $\bar{K^\circ}$, gives 
the most stringent bound on the background torsion  to be
$\langle T^0 \rangle < 10^{-25}$ GeV.

\vskip .3in
{\bf Acknowledgment}

One of us (US) would like to thank the Abdus Salam International 
Center for Theoretical Physics for hospitality during his visit 
as an Associate, where this work has been completed.

\end{document}